\documentclass[12pt,A4paper]{article}
\pdfoutput=1

\usepackage{indentfirst} 
\usepackage{cite}
\usepackage{graphicx,epstopdf}

\def\br{{\cal B}}
\def\gam{\gamma}
\def\beq{\begin{equation}}
\def\eeq{\end{equation}}
\def\fbi{~{\rm fb}^{-1}}

\begin{document}

\setcounter{page}{0}\thispagestyle{empty}

\begin{center}

\vspace*{0.5cm}

{\Large\bf On the presentation of the LHC Higgs Results}

\vspace*{1cm}

{\em Conclusions and suggestions from the workshops \\ 
``Likelihoods for the LHC Searches'', 21--23 January 2013 at CERN, \\
``Implications of the 125~GeV Higgs Boson'', 18--22 March 2013 at LPSC Grenoble, \\  
and from the 2013 Les Houches ``Physics at TeV Colliders'' workshop.}

\vspace*{1cm}

F.~Boudjema,$^{a}$ 
G.~Cacciapaglia,$^{b}$   
K.~Cranmer,$^{c}$ 
G.~Dissertori,$^{d}$  
A.~Deandrea,$^{b}$   
G.~Drieu la Rochelle,$^{b}$ 
B.~Dumont,$^{e}$ 
U.~Ellwanger,$^{f}$ 
A.~Falkowski,$^{f}$  
J.~Galloway,$^{g}$   
R.\,M. Godbole,$^{h}$  
J.\,F.~Gunion,$^{i}$ 
A.~Korytov,$^{j}$ 
S.~Kraml,$^{e}$ 
H.\,B.~Prosper,$^{k}$ 
V.~Sanz,$^{l}$ 
S.~Sekmen$^{m}$

\vspace*{1cm}

{\small \it 
$^a\,$LAPTH, Universit\'e de Savoie, CNRS, B.P.110, F-74941 Annecy-le-Vieux Cedex, France \\[1mm]
$^b\,$Universit\'e de Lyon, F-69622 Lyon, France, Universit\'e Lyon 1, CNRS/IN2P3, UMR5822 IPNL, F-69622 Villeurbanne Cedex, France\\[1mm]
$^c\,$Department of Physics, New York University, New York, NY 10003, USA\\[1mm]
$^d\,$Institute for Particle Physics, ETH Zurich, Schafmattstrasse 20, HPK E28, CH-8093 Zurich, Switzerland\\[1mm]
$^e\,$Laboratoire de Physique Subatomique et de Cosmologie, UJF Grenoble 1,
CNRS/IN2P3, INPG,\\ 53 Avenue des Martyrs, F-38026 Grenoble, France\\[1mm]
$^f\,$Laboratoire de Physique Th\'eorique, UMR 8627, CNRS and
Universit\'e de Paris--Sud, F-91405 Orsay, France\\[1mm]
$^g\,$Dipartimento di Fisica, Universit\`a di Roma ``La Sapienza"  and INFN Sezione di Roma, I-00185 Roma, Italy\\[1mm]
$^h\,$Centre for High Energy Physics, Indian Institute of Science, Bangalore 560012, India\\[1mm]
$^i\,$Department of Physics, University of California, Davis, CA 95616, USA\\[1mm]
$^j\,$Department of Physics, University of Florida, Gainesville, FL 32611, USA\\[1mm]
$^k\,$Department of Physics, Florida State University, Tallahassee, FL 32306, USA\\[1mm]
$^l\,$Department of Physics and Astronomy, University of Sussex, Brighton BN1 9QH, UK\\[1mm]
$^m\,$CERN, CH-1211 Geneva 23, Switzerland
}

\vspace*{1cm}
\abstract{We put forth conclusions and suggestions regarding the presentation of the LHC Higgs results 
that may help to maximize their impact and their utility to the whole High Energy Physics community.}

\end{center}

\clearpage 
\section{Motivation} 

The LHC was built to explore the TeV energy scale in order to unravel
the mechanism of electroweak symmetry breaking~(EWSB) and shed light on 
physics beyond the Standard Model (SM) of electroweak and strong
interactions. 
The discovery~\cite{atlas:2012gk, cms:2012gu} in July 2012 of a new particle with mass around 125~GeV 
and with properties consistent with those of a SM Higgs boson is thus a first triumph for the LHC
physics program. 

However, while this discovery  completes our picture of the SM, it still 
leaves many fundamental questions open. 
In particular, the SM does not explain the value of the electroweak~(EW) scale itself. 
And, there is the closely related issue of understanding why the Higgs boson is so light.  
In the absence of New Physics, both the Higgs mass and the EW scale are predicted to be driven to the scale of Grand Unification~($M_{\rm GUT}$), or even the Planck scale, by radiative corrections.
The New Physics at the electroweak scale needed to avoid this inevitably implies that the properties ({\it i.e.}\ the couplings) of the Higgs boson primarily associated with electroweak symmetry will differ from SM predictions. Therefore, a prime goal for the near future is to look for deviations of  this Higgs-like signal from the SM predictions. 

Indeed, with the 25~fb$^{-1}$ of data collected at $\sqrt{s} = 7$ and 8 TeV, the analyses of the ATLAS and CMS 
collaborations are beginning to provide a comprehensive picture of the production and decay properties 
of the 125 GeV Higgs boson~\cite{ATLAS:2013comb,Aad:2013wqa,CMS:2013comb}. In particular,
having detailed measurements in various  production/decay channels greatly increases the potential for revealing anomalies relative to the SM.  In fact, given the absence of direct signals for New Physics beyond the SM (BSM) at the LHC, the Higgs data and their interpretation currently provide the crucial guidelines for BSM theories. Consequently, in-depth studies of the Higgs signal could have profound implications for supersymmetric models, Randall-Sundrum models (with Higgs-radion mixing), technicolor models, little Higgs models, composite Higgs models, non-minimal Higgs sectors and so on.

Fits to various combinations of reduced Higgs couplings ({\it i.e.} Higgs couplings to fermions and gauge bosons relative to their SM values) have been performed by the experimental collaborations themselves~\cite{ATLAS:2013comb,CMS:2013comb} and this work will surely continue. However, given the variety of BSM models, their variants, and the near certainty that their numbers will grow, it is crucial that the experimental collaborations present results in a way that maximizes their utility to the broader scientific community. 
There has indeed been a boom 
of phenomenological papers making use of the published Higgs results and investigating their implications for BSM physics. For example,  there is a large effort ongoing 
to study the implications of the LHC Higgs results in an effective Lagrangian approach, see~\cite{Carmi:2012yp,Azatov:2012bz,Espinosa:2012ir,Klute:2012pu,Azatov:2012wq,Low:2012rj,Corbett:2012dm,Giardino:2012dp,Ellis:2012hz,Montull:2012ik,Espinosa:2012im,Carmi:2012in,Banerjee:2012xc,Bertolini:2012gu,Bonnet:2012nm,Plehn:2012iz,Elander:2012fk,Djouadi:2012rh,Dobrescu:2012td,Moreau:2012da,Cacciapaglia:2012wb,Corbett:2012ja,Masso:2012eq,Azatov:2012qz,Belanger:2012gc,Cheung:2013kla,Celis:2013rcs,Belanger:2013kya,Falkowski:2013dza,Cao:2013wqa,Giardino:2013bma,Ellis:2013lra,Djouadi:2013qya,Chang:2013cia,Dumont:2013wma,Bechtle:2013xfa,Belanger:2013xza}. Likewise, many studies, 
which we do not have space to cite here, are performed for specific models such as those mentioned above. 
Clearly, it will be of great value if the experimental results are presented in a way that obviates the (current) need by the broader community to make unnecessarily crude approximations in such studies. 

In this document, we therefore advocate a systematic way of presenting the LHC Higgs results. 
In doing so, we build upon the  recommendations given in the  
``Les Houches Recommendations for the Presentation of LHC Results''~\cite{Kraml:2012sg}, which  
stressed the importance of providing all relevant information, including the best-fit signal strengths, $\mu$, on a channel-by-channel basis for the independent production and decay processes. 
``Interim recommendations to explore the coupling structure of a Higgs-like particle,'' including detailed discussions of coupling scale factors and benchmark parametrizations, were given by the LHC Higgs Cross Section Working Group in \cite{LHCHiggsCrossSectionWorkingGroup:2012nn}. 
The purpose of the present document is to discuss in more detail issues involved in the interpretation in general BSM models and how to make the Higgs results maximally usable for the whole high-energy-physics community. 
In order to motivate our recommendations and the need for them, we discuss the problems that the use of incomplete experimental information can cause and their potential impact  on the community's ability to navigate systematically  a rapidly expanding network of experimental and theoretical results.  
We then make a concrete set of proposals  for a coherent and systematic approach to the release of Higgs sector results, which, if adopted by all experimental groups,  could yield significant benefits to the field. 
 
In theories beyond the SM,  the Higgs production cross sections, decay branching ratios, kinematic distributions, and even the number of Higgs particles may differ from SM predictions. 
It is helpful to make a clear distinction between two classes of models distinguished by whether or not  the selection efficiency and detector acceptances for various interactions are independent of the model parameters; in both cases, the cross sections and branching ratios may vary. The former case is considerably easier as the parametrization of the likelihood comes from a simple scaling, while the latter case is difficult, because the efficiencies and acceptances can depend non-trivially on the model parameters.  Sections~2, 3 and 5 focus on the former case in which the tensor structure of the Higgs-like particle is specified and universal efficiencies can be calculated for individual processes.  Sections~\ref{sec:tensor} and \ref{sec:fiducial} focuses on a strategy for a more generic class of models in which the efficiencies and acceptances can vary.

\section{Disentangling multiple production modes} \label{sec:muXY}

In this section we focus on presenting results in theories with SM-like tensor structure, in which the efficiency and acceptances are constant with respect to the parameters of the theory. Considerable progress can be
made in studying these models given the proper information.
In particular, a detailed breakdown in terms of production mode and decay is 
needed for testing such models. 
Below we use $X$ to denote the fundamental production mechanisms, such as gluon fusion, $gg\to H$  (ggF),  
vector-boson fusion, $WW\to H$ and $ZZ\to H$ (VBF), ZH, WH, or ttH associated production; 
and $Y$ to denote the Higgs decay final states ($Y=\gam\gam$, $WW$, $ZZ$, $b\bar b$ and $\tau\tau$ are currently accessible). 

Initially, the most common representation of LHC Higgs results was in terms of  the \textit{global signal strength}, $\mu$, which scales all pairs $(X,Y)$ simultaneously according to
\begin{equation}
    {\sigma(X)\br(H\to Y)} = \mu \; \sigma(X_{\rm SM} )\br(H_{\rm SM}\to Y)\,,
\end{equation}
with $\sigma$ the $pp$ production cross section  for a given Higgs production mode and $\br$ the decay branching ratio. 
This global signal strength ties together multiple production modes to their SM ratios, which is almost never
a property of theories beyond the SM.
Instead, for each $(X,Y)$ pair, we can define the scale factor $\mu(X,Y)$ with respect to the SM Higgs:
\begin{equation}
   \mu(X,Y)\equiv {\sigma(X)\br(H\to Y)\over \sigma(X_{\rm SM} )\br(H_{\rm SM}\to Y)}\,.
\end{equation}
The likelihood in terms of these $\mu(X,Y)$ allows for reinterpretation of the results within the class of models where the efficiency and acceptance for each $(X,Y)$ pair are approximately unchanged with respect to the SM.\footnote{Note here that the determination of signal strength modifiers $\mu$ entangles a variety of assumptions which are based on the SM Higgs hypothesis. These include, {\it e.g.}, the narrow width approximation, assumptions on relative cross sections for different production mechanisms, theoretical uncertainties on these cross sections, assumptions on
efficiencies/acceptances (in particular for 0-jet, 1-jet, and 2-jet selections), the assumption that there are no new production mechanisms, and even assumptions on the choice of probability density
functions to be associated with the theoretical uncertainties.}

In practice,  the data related to a single decay mode 
$H \rightarrow Y$ are divided into different categories (or ``sub-channels'') $I$,  in order to improve 
sensitivity or discrimination among the production mechanisms $X$.  As an example, for the $\gam\gam$ final state the categories include ``untagged'', 2-jet tagged, and lepton tagged categories, designed to be most sensitive to ggF, VBF, and VH, respectively.  We denote the global signal strength for a specific category by $\mu_I(Y)$.

\begin{figure}[t]
\hspace*{-8mm}
		\includegraphics[width=6.2cm]{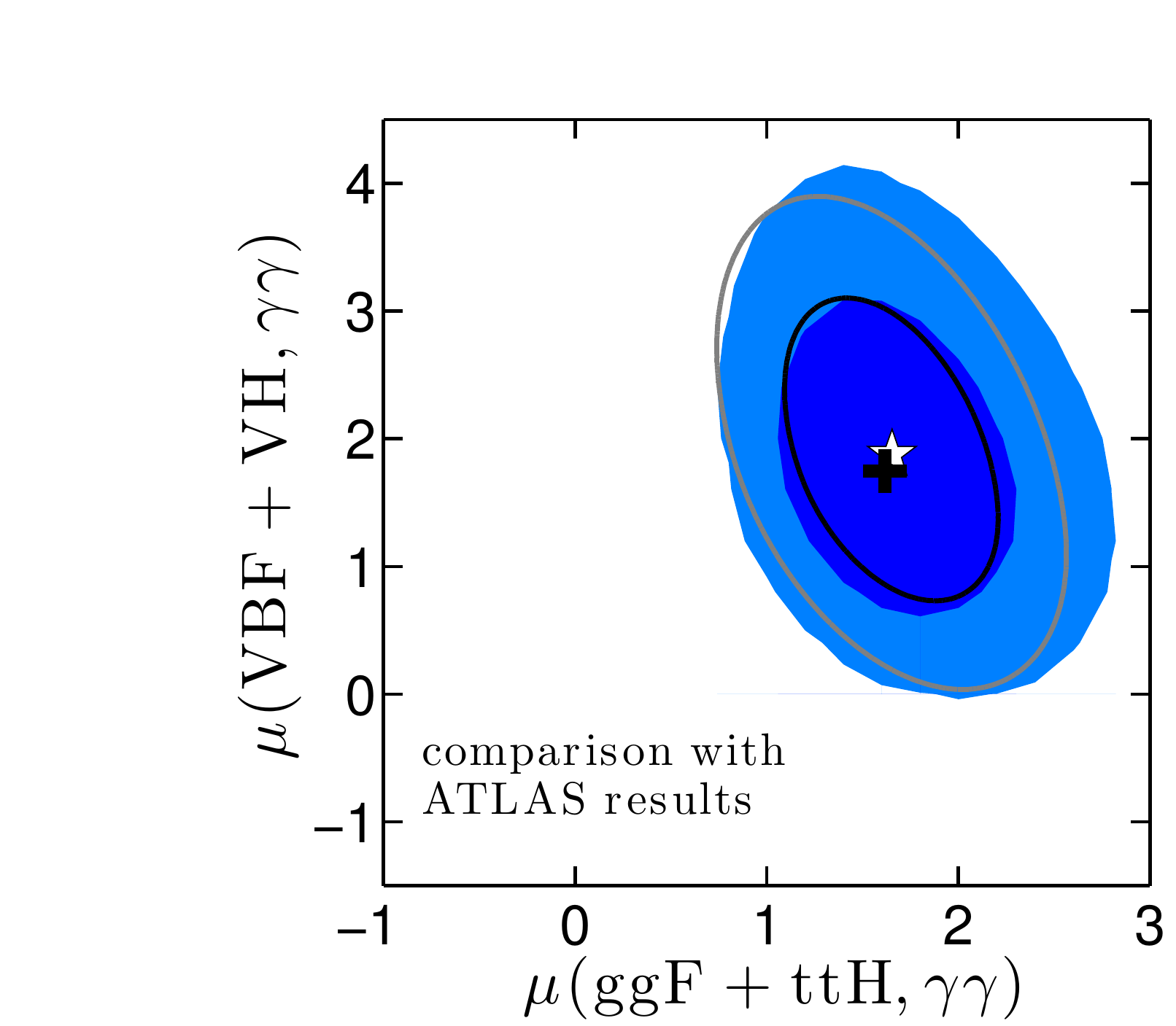}\quad
		\includegraphics[width=5cm]{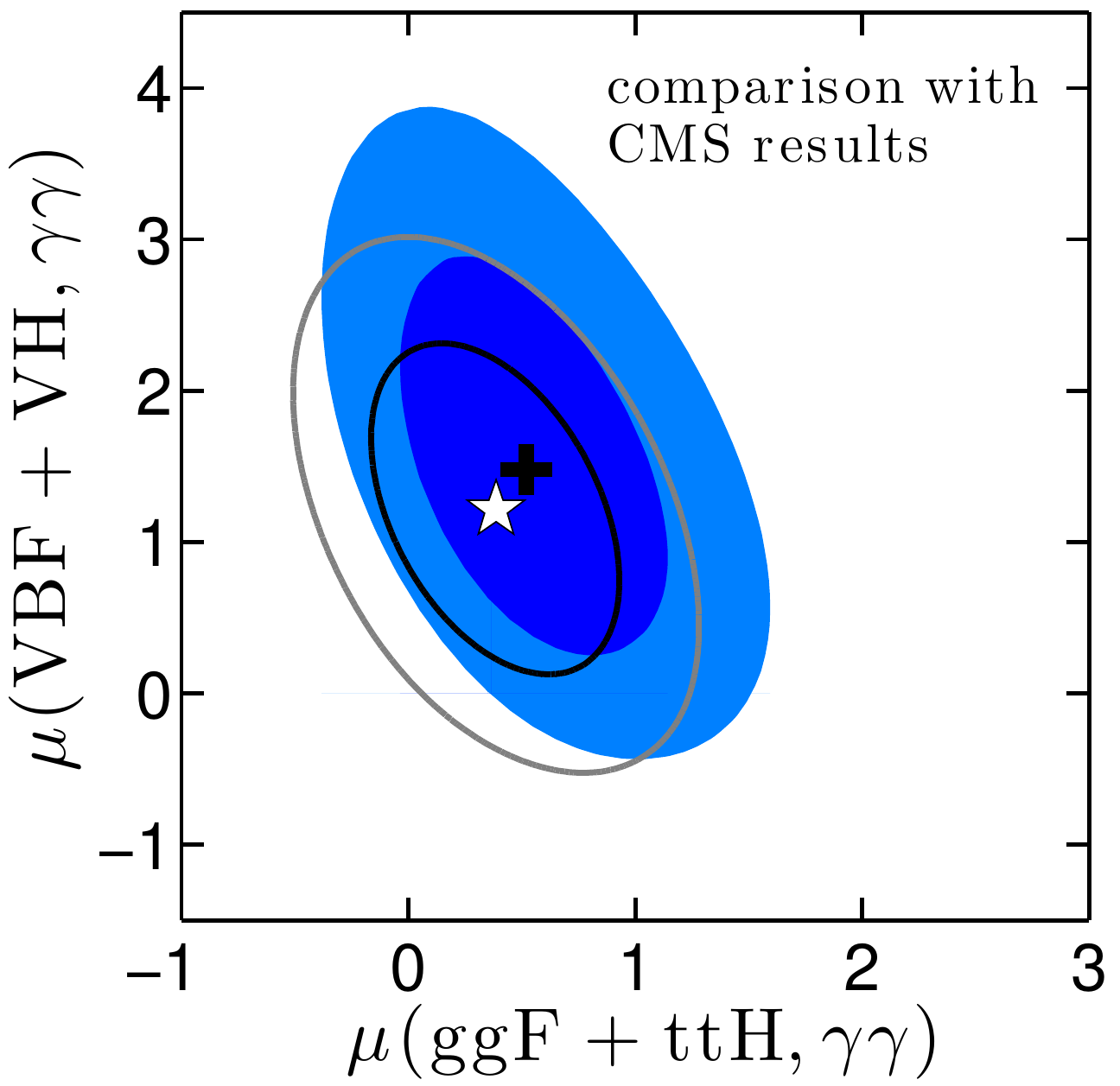}\quad
		\includegraphics[width=5cm]{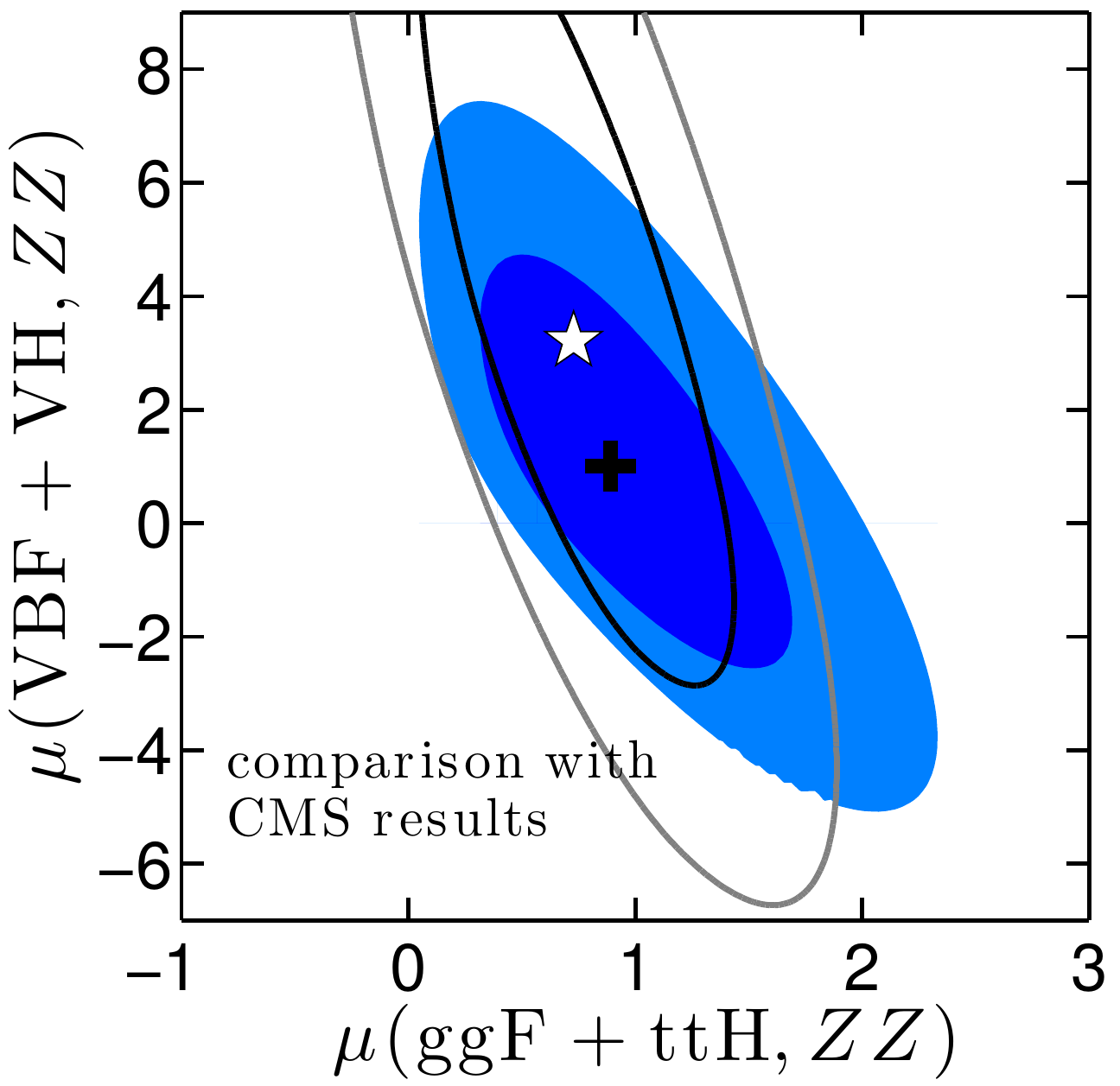}
	\caption{Reconstructing the likelihood from subchannel information. The black and gray lines show the 68\% and 95\%~CL contours in the $(\mu_{\rm ggF + ttH}, \mu_{\rm VBF + VH})$ plane, reconstructed from signal strengths and efficiencies for the experimental categories $I$ in each final state; from left to right for  the ATLAS $H \rightarrow \gamma\gamma$~\cite{ATLAS-CONF-2013-012}, CMS $H \rightarrow \gamma\gamma$ (MVA)~\cite{CMS-PAS-HIG-13-001}, and CMS $H \rightarrow ZZ$~\cite{CMS-PAS-HIG-13-002} channels.  For comparison, the dark and light blue filled areas show the 68\% and 95\%~CL regions directly given by the collaborations. The black crosses are the experimental best fit points,  the white stars are the reconstructed ones. 
\label{fig:reconstruction}}
\end{figure}

Although the categories $I$ are typically designed using cuts and/or tags that maximize sensitivity to the known fundamental production mechanisms $X$, cuts cannot be designed so that a given $I$ is sensitive to only one $X$.  
It is critical that  for each of the categories $I$ the total selection efficiency (including detector acceptance)  be provided for each production mode.   
As demonstrated in Figure~\ref{fig:reconstruction}, 
the likelihood function in terms of $\mu(X,Y)$  can be approximately recomputed combining the $\chi^2$ of all categories $I$ using an efficiency-weighted sum to match the overall signal strength,
\beq
\mu_I (Y)= \sum_X   \mu(X,Y) \,   T(I,X) \, \sigma(X_{\rm SM}) \br(H_{\rm SM}\to Y)\,,
\eeq
where $T(I,X)$ are the selection efficiencies for each production mode, normalized to 1.
This issue was noted some time ago, and the ATLAS and CMS Collaborations have progressively expanded their release of this information; however, the process of providing the complete information is still a work in progress (for example, it is not yet available for the $Y=WW$ or $ZZ$ channels in ATLAS). 

However, even with the full knowledge of the efficiencies $T(X,I)$, this approach is limited and may lead to 
partly unreliable results because  
important correlations may be missed. In particular, some systematic uncertainties lead to migration of events between categories, and these uncertainties can dominate over the  statistical ones. 

Instead of extracting the $\mu(X,Y)$ from the global signal strengths $\mu_I$, a more direct approach is  for the experimental collaborations to explicitly present, for each decay mode $Y$, the full experimental likelihood as a function of multiple production modes.  
At this moment in time it is convenient and relevant (see later discussion) to combine the 5 usual production modes ($X$ = ggF, ttH, VBF, ZH, WH)  to form just two effective $X$ modes (${\rm ggF+ttH}$ and ${\rm VBF+VH}$, where ${\rm VH} = {\rm ZH}+{\rm WH}$). The likelihood can then be shown in the $(\mu_{\rm ggF + ttH},\mu_{\rm VBF + VH})$ plane for each final state. 
Such figures are progressively becoming a standard, see {\it e.g.}~\cite{Aad:2013wqa, ATLAS:2013comb,CMS:2013comb}. 

The information in the $(\mu_{\rm ggF + ttH}, \mu_{\rm VBF + VH})$ plane is a boon for interpretation studies for two reasons. 
First, in presenting such results the experimental collaborations have effectively unfolded the $I$ contributions to each of the two $X$ channels  listed above. In other words, due to the fact that the production mechanisms are properly taken into account, there is  no need to know efficiencies and best fits in the individual $I$ categories and  correlations.  Second, an approximation to the full likelihood---assuming no correlation between the various final states---can be derived from a single plot (see {\it e.g.}\ Figure~7 from~\cite{Aad:2013wqa} and Figure~4 from~\cite{CMS:2013comb}) by fitting the 68\% or 95\% CL contours with a 2D normal distribution. 
This approach is straightforwardly generalized to the $(\mu(X,Y), \mu(X',Y))$ plane, where the $X$ and $X'$ are appropriate for the analysis at hand.  For example, the $H\to b\bar{b}$ analyses naturally probe WH and ZH production.  Moreover, a three-dimensional scan over ggF+ttH, VBF, WH should be feasible and appropriate for $H\to\gam\gam$.  
 
\begin{figure}[t]
	\centering
		\includegraphics[width=5cm]{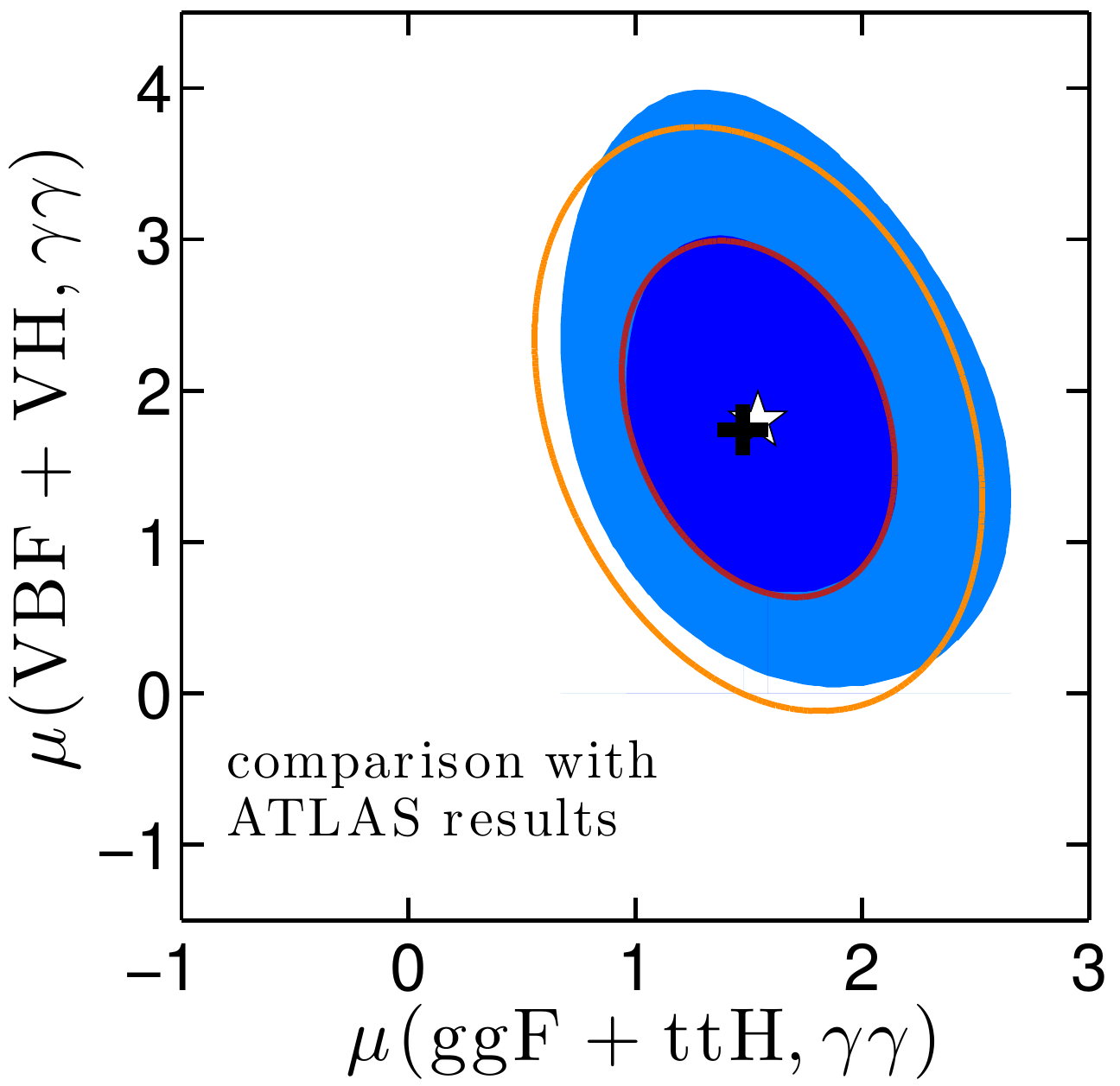}\quad
		\includegraphics[width=5cm]{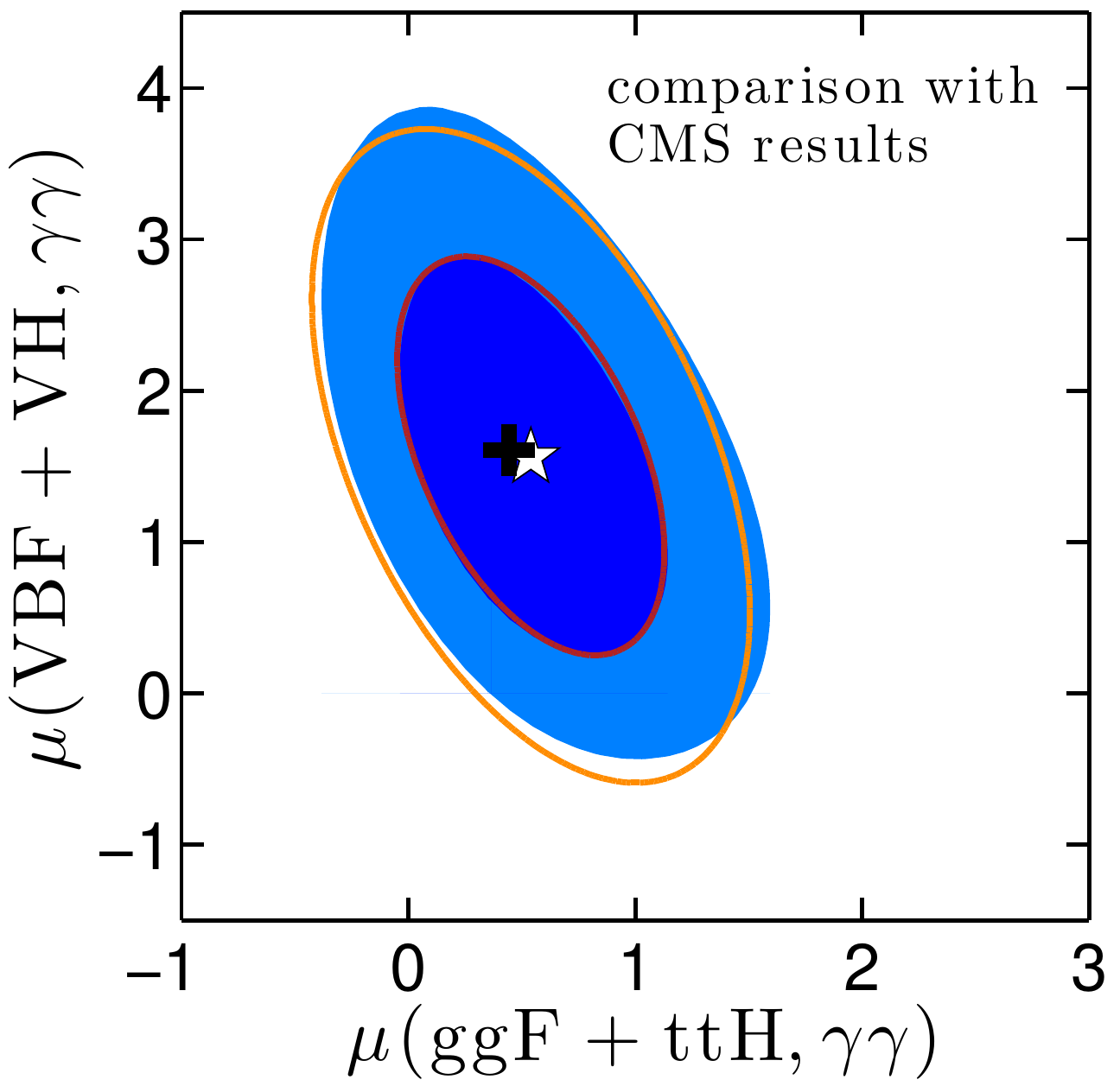}\quad
		\includegraphics[width=4.9cm]{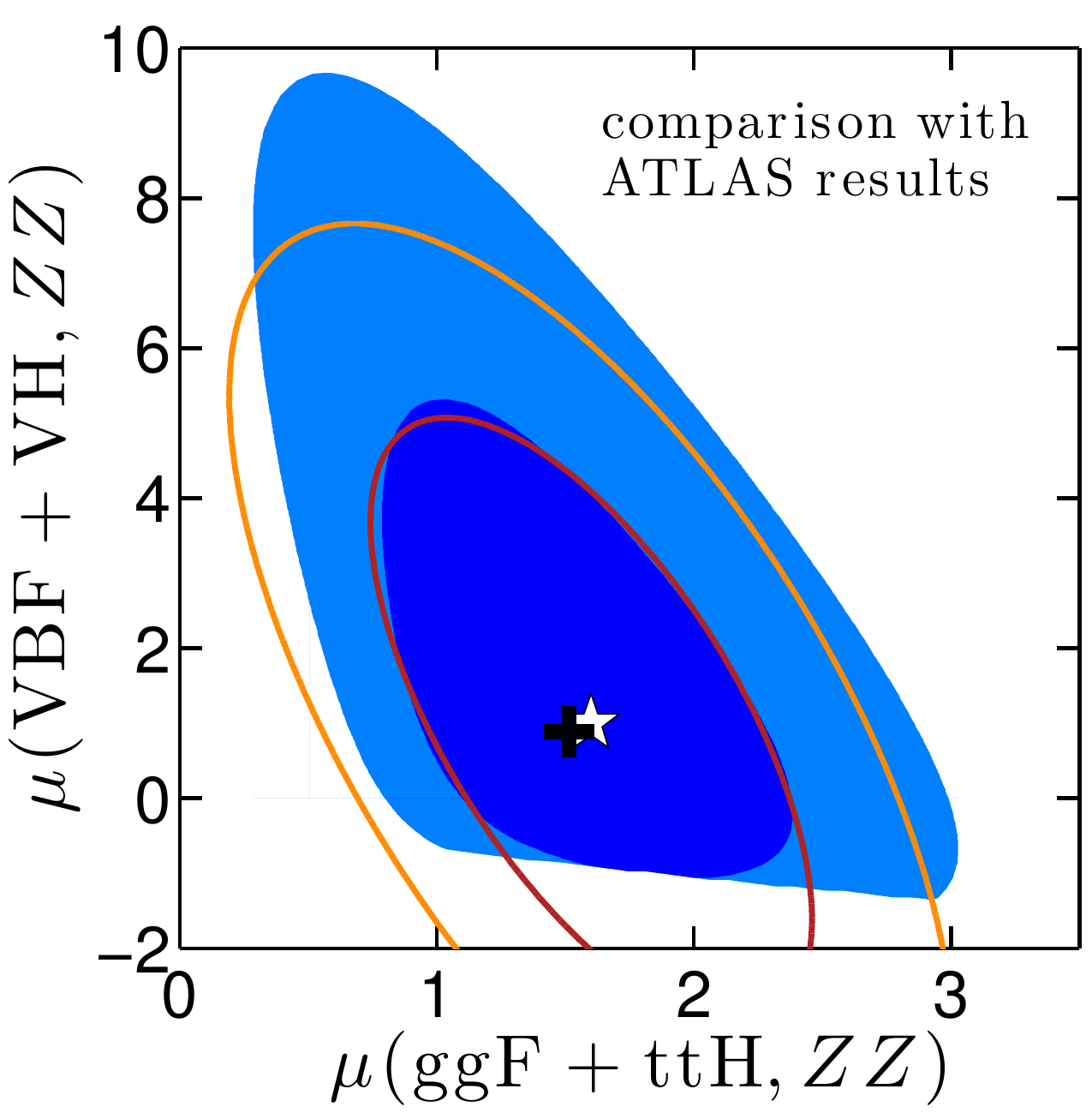}
	\caption{Gaussian fit to signal strenghts in the $(\mu_{\rm ggF + ttH}, \mu_{\rm VBF + VH})$ plane, from left to right for  the ATLAS $H \rightarrow \gamma\gamma$~\cite{Aad:2013wqa}, CMS $H \rightarrow \gamma\gamma$ (MVA)~\cite{CMS-PAS-HIG-13-001}, and ATLAS $H \rightarrow ZZ$~\cite{Aad:2013wqa} channels. The dark and light blue filled areas are the 68\% and 95\%~CL regions given by the experiments, the red and orange lines show the fitted ones. In all three cases, we approximately reconstruct the likelihood by fitting a bivariate normal distribution to the 68\% CL contour given by the collaboration (see also \cite{Cacciapaglia:2012wb,Belanger:2012gc,Belanger:2013xza}). The black crosses are the experimental best fit points, while the white stars are the mean values from the fit. 
\label{fig:GaussianFit}}
\end{figure}

\begin{figure}
	\centering
		\includegraphics[width=0.4\textwidth]{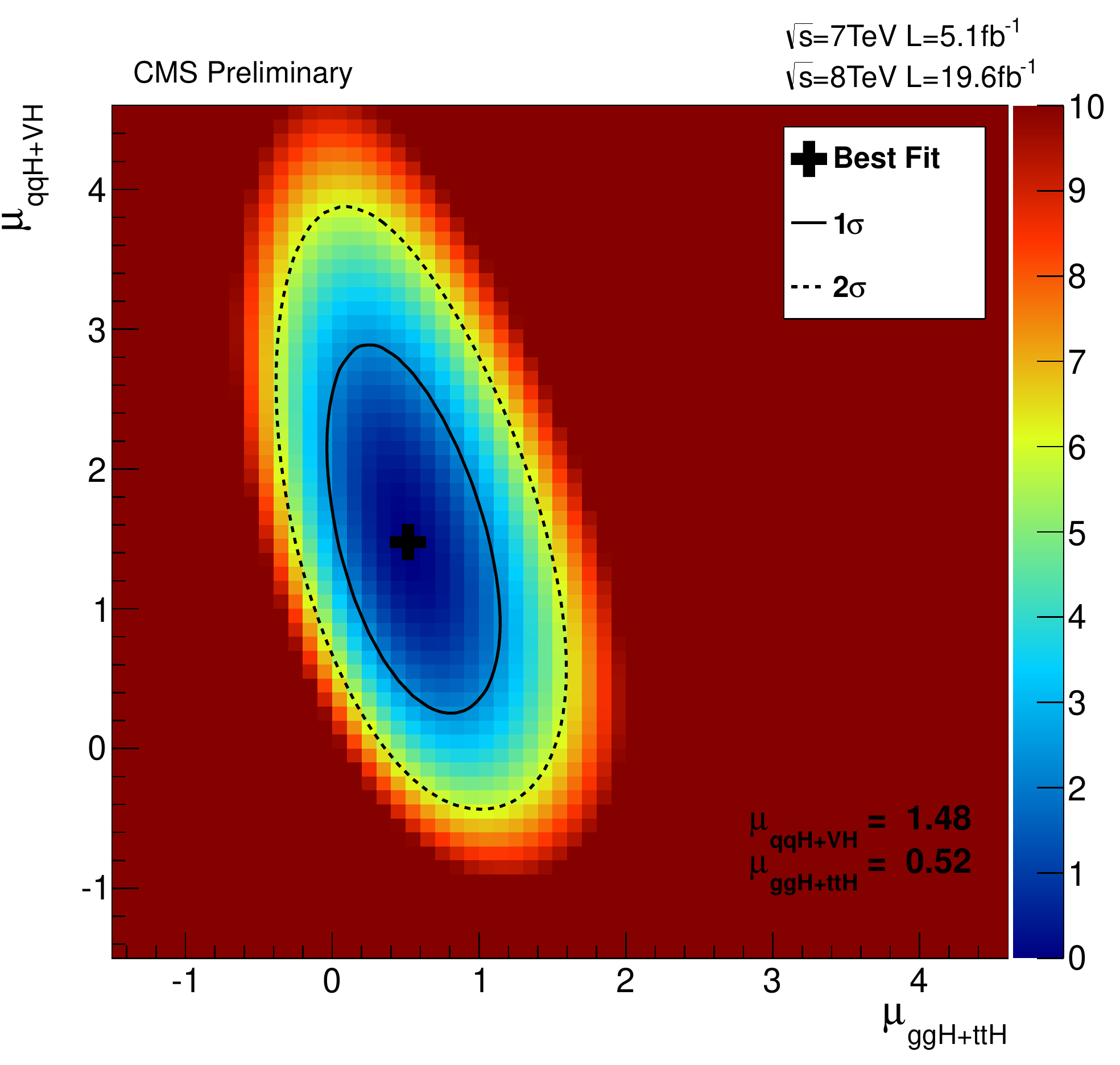}	
\caption{An example from CMS~\cite{CMStwiki} of the likelihood in the $\mu({\rm ggF+ttH}, \gam\gam) - \mu({\rm VBF+VH}, \gam\gam)$ plane.  The color indicates the value of the likelihood, which conveys more information than just the contours.  Preferably this information would be directly available in numerical form via INSPIRE~\cite{inspire}.  See Appendix~\ref{A:appendix}.
\label{fig:CMS_gamgam_like}}
\end{figure}

A practical difficulty that can trivially be avoided stems from the fact  that typically only the 68\% and 95\% CL contours are displayed in the $\mu_{\rm ggF + ttH}$ versus $\mu_{\rm VBF + VH}$ plots.  
In order to use this information, one thus first needs to reconstruct the likelihood function. The simplest assumption---having normally distributed signal strengths---is not fully satisfactory and sometimes reproduces the contours rather poorly, as in ATLAS $H \rightarrow ZZ$. 
This is illustrated in Figure~\ref{fig:GaussianFit}, which compares the experimental contours with the Gaussian approximation fitted from the 68\% contour, for ATLAS and CMS $H \rightarrow \gamma\gamma$, and for ATLAS $H \rightarrow ZZ$. It would be of great advantage to have the full likelihood information in the $(\mu_{\rm ggF + ttH}, \mu_{\rm VBF + VH})$ plane. 
The CMS collaboration already provided this for the $H \rightarrow \gamma\gamma$ analysis as supplementary material on their TWiki web page~\cite{CMStwiki},  shown  here in Figure~\ref{fig:CMS_gamgam_like}. 
It would be extremely helpful if such ``temperature'' plots were adopted as the standard and
provided in numerical form (for instance, by providing the numerical content of the likelihood plots {\it i.e.} the likelihoods, over a grid on the $(\mu_{\rm ggF + ttH}, \mu_{\rm VBF + VH})$ plane in a table, or providing the plots in an electronic form such as a {\tt ROOT} file, etc.).\footnote{Indeed, soon after the first version of this note 
was published, the ATLAS collaboration already made a first step in this direction, see Appendix~\ref{A:appendix}.}

Note finally that ratios of $\mu$'s at different energies provide important information on anomalous couplings, as they are sensitive to different momentum dependence. It would thus be valuable to eventually have ratios of signal strengths from different runs ($\sqrt{s}=7,\, 8,\, 13,\, 14$~TeV), where the correlations on systematic errors are taken into account.

\section{Towards the full likelihood information} \label{sec:beyond}

Grouping together VBF+VH, {\it i.e.}\ rescaling the VBF, WH and ZH production mechanisms 
by a common factor, 
is justified theoretically in models with custodial symmetry, while grouping together ggF+ttH is well justified 
given the current level of precision in probing ttH associated production. 
Eventually, however, we want to test ggF, ttH, VBF, ZH and WH separately, which means that we need 
a more detailed break down of the channels beyond 2D plots.

The optimum would of course be to have the full statistical model 
available,  and methods and tools are indeed being developed~\cite{publishL} to make this feasible,  
{\it e.g.}, in the form of {\tt RooFit} workspaces. 
However, it may still take a while until likelihoods will indeed be published in this way.  
We would therefore like to advocate as a compromise that the experiments give the likelihood for each final state $Y$ as a function of a full set of production modes, that is to say, in the 
\beq
 (m_H, \mu_{\rm ggF},\mu_{\rm ttH},\mu_{\rm VBF},\mu_{\rm ZH},\mu_{\rm WH}) 
\eeq
parameter space. By getting the likelihood function in this form for each decay mode, a significant step could be taken towards a more precise fit in the context of a given BSM theory.
Note that the signal strengths' dependence on $m_H$ is especially important  
for the high-resolution channels ($\gamma\gamma$ and $ZZ$, also $Z\gamma$ in the future). While the signal strengths seem to form a plateau in the case of $H \rightarrow \gamma\gamma$ (at least in ATLAS), there is a very sizable change in the $H \rightarrow ZZ$ channel if we change $m_H$ by 1 or 2 GeV.

The likelihood could be communicated either as a standalone computer library or as a large grid data file. This choice is mostly meant to be an intermediate step between a full effective Lagrangian parameterization (which would be difficult to communicate) and simple 2D parameterizations which unfortunately do not cover all the theoretical possibilities. Having the full likelihood shape and not just some contours would allow the community to overcome the Gaussian limitation.

\section{Tensor structure of Higgs boson couplings} \label{sec:tensor}

Apart from the Higgs production and decay rates, 
experiments can probe differential distributions of decay products in Higgs $n$-body decays with $n>2$  
which carry valuable information about the tensor structure of the Higgs couplings. 
For example, in the case of $H \to V V^* \to 4f$ decays (assuming massless fermions),  
the Higgs boson $H$ couplings to the SM gauge bosons can be parametrized as 
\begin{equation}
\label{eq:SpinZeroAmplitude} 
{\cal A} (H \to V_{\mu}^1  V_{\nu}^2 ) = {1 \over v} \left (
F_1(p_1^2,p_2^2) 2 m_V^2 \eta_{\mu \nu}  +  F_2(p_1^2,p_2^2) p_{1\nu} p_{2\mu} +   F_3(p_1^2,p_2^2)  \epsilon_{\mu \nu  \rho \sigma}  p^\rho_1 p^\sigma_2
\right ) \,, 
\end{equation}
with some form factors $F_{1,2,3}$.
At the zeroth order in the vector boson momentum expansion,  
the first form factor is a constant, $F_1 = a_1$ and $F_{2,3} = 0$.
Note that $(F_1,F_2,F_3)=(1,0,0)$ correspond to the SM Higgs boson at the tree level.

However, the LHC experiments can already probe the presence of non-zero ${\cal O}(p^2)$ terms
with the caveat that the SM loop-induced contributions to these terms 
are not measurable even at the ultimate luminosities of the HL-LHC. 
At that order, one should consistently take into account 
both the leading order terms of the $F_{2,3}$ form factors, $F_{2,3} = a_{2,3}$ (constants), 
as well as the next-to-leading term in the momentum expansion of $F_1$,  
$F_1 =  a_1+ a_4(p_1^2 + p_2^2)$.

For $H \to ZZ^*$ decays, the Higgs couplings to two $Z$ bosons that can arise from up to dimension-6 operators in  the effective Higgs Lagrangian are given by \cite{Contino:2013kra}:
\begin{equation}
\label{eq:SpinZeroLagrangian}
\mathcal{L} \supset H 
\left[ \, \kappa_1 \frac{m_Z^2}{v} Z_\mu Z^\mu +
\frac{\kappa_2}{2 v} Z_{\mu\nu} Z^{\mu\nu} +
\frac{\kappa_3}{2 v} Z_{\mu\nu}\tilde{Z}^{\mu\nu}
 + \frac{\kappa_4}{v}  Z_\mu \partial_\nu Z^{\mu \nu} \,
\right].
\label{lagrangian}
\end{equation}
where $Z_{\mu \nu} = \partial_\mu Z_\nu -  \partial_\nu Z_\mu$, and $\tilde{Z}_{\mu\nu} = \frac{1}{2}\epsilon_{\mu\nu\rho\sigma} Z^{\rho \sigma}$.
Couplings $(\kappa_1,\kappa_2,\kappa_3)=(1,0,0)$ correspond to the SM Higgs boson.
The decay amplitude form factor constants $a_i$ in Eq.~(\ref{eq:SpinZeroAmplitude}) 
and the effective Lagrangian couplings $\kappa_i$ in Eq.~(\ref{eq:SpinZeroAmplitude}) 
are related to each other by 
$(a_1, a_2, a_3, a_4) = (\kappa_1 - \kappa_2 \frac{m_H^2}{2m_Z^2}, 2\kappa_2, 2\kappa_3, (\kappa_2+\kappa_4) \frac{1}{2m_Z^2} )$.

It would be enlightening if the experimental Higgs boson results in the $H\to VV^*$ decays channels 
could be recast as constraints on the $(a_1, a_2, a_3, a_4)$ or $(\kappa_1, \kappa_2, \kappa_3,\kappa_4)$ parameters.
At present,  this is only partially borne out and only in the $H \to ZZ^* \to 4l$ channel, 
where the likelihood of  $f_{a_3} \equiv  \sigma_3/(\sigma_1 + \sigma_3)$ was presented by CMS~\cite{CMS-PAS-HIG-13-002}.
Here, $\sigma_1$ and $\sigma_3$ are $4\ell$ cross sections corresponding
to the $a_1$- and $a_3$-terms in the $H \to ZZ$ decay amplitude form factors,
or the $\kappa_1$- and $\kappa_3$-terms in the Lagrangian given by Eq.~(\ref{eq:SpinZeroLagrangian}).
In this measurement, $\kappa_2$ and $\kappa_4$ are assumed to be zero and, hence, so are $a_2$ and $a_4$.
If results are presented in this form, {\it i.e.} in the form of fractional cross sections, 
experiments must be very clear whether cross sections $\sigma_i$ are in the fiducial region 
or efficiency-corrected total cross sections. 
For the same set of couplings $(\kappa_1, \kappa_2, \kappa_3,\kappa_4)$,
ratios of cross sections for same-flavor leptons, {\it e.g.} $\sigma_i (4\mu) / \sigma_j (4\mu)$, 
different-flavor leptons, $\sigma_i (2e2\mu) / \sigma_j (2e2\mu)$, 
all $e/\mu$-leptons, $\sigma_i (4\ell) / \sigma_j (4\ell)$, etc.\ can differ by as much as
$\sim$10--20\%, which stems directly from the interference effects associated with
permutations of identical leptons in the final states. 
Clear definitions would allow for an unambiguous translation of experimental results 
expressed as fractional cross sections into limits on or measurements of effective couplings.

The total cross section is
affected by the couplings on the production side of the new boson; hence, absolute
values of couplings associated with $H \to V V^* \to 4f$ decays cannot be unraveled
from studying a single decay channel. 
With the total cross section constrained by data, the spin-parity analysis has $(n-1)$ degrees of freedom,  where $n=4$  in the example above  (assuming that all constants, $a_i$ and $\kappa_i$, are real).
Therefore, ideally, the experiments should present the spin-parity
results as a $(n-1)$-dimensional likelihood, {\it e.g.}, 
$\mathcal{L}(\kappa_2/\kappa_1, \kappa_3/\kappa_1, \kappa_4/\kappa_1)$.  
Such likelihoods would allow theorists to place robust limits on a large class of scenarios beyond the Standard Model.

Note that, while we do not discuss this option here, complex form factors can in principle be treated in an analogous way. 
Finally, we note that also differential distributions of the associated jets in VBF~\cite{Englert:2012xt,Djouadi:2013yb}, 
as well as  polarization and kinematic distributions of the vector bosons in VH~\cite{Ellis:2013ywa,Godbole:2013saa}  production, carry important information 
and can help probing the structure of the $HWW$ vertex separately from the $HZZ$ vertex. 

\section{Results for additional Higgs-like states}\label{sec:additional}

Searching for additional Higgs-like states $\phi$ with masses
above or below 125 GeV is interesting and well motivated.  
For additional states having the same production and decay channels 
and tensor structures for the couplings as the SM Higgs, 
we advocate that the observed and expected limits on signal rates be 
shown systematically as functions of $M_\phi$, including the injection of a
SM-like Higgs boson at 125--126 GeV. 
This has been done already by CMS in the $H\to\gam\gam$ analysis, see Figures~2 and 3 of \cite{CMS-PAS-HIG-13-016}.

Indeed, the contributions from a SM-like Higgs boson at 125--126
GeV can---and should---be treated like any other SM background, and
the limits on signal rates of additional Higgs-like bosons can be shown once the
contributions from a SM-like Higgs boson at 125--126 GeV to a given
search channel are subtracted. This
injection is especially important when looking for excesses in search
channels with
low mass resolution ($b\bar b$, $WW$, $\tau\tau$).

The injection of the Higgs boson at 125--126~GeV as additional background 
could be based on the expectations for a 100\% SM-like
Higgs boson, or on the best fit to the observed signal rates. The latter
would depend on the amount of data accumulated, which channels (and
which experiments) are combined and, in any case, be subject to
systematic and statistical errors. To be conservative we propose to use
the well-defined contributions from a SM-like Higgs boson to this end.
(If in the future there are substantial deviations of the observed
125--126 GeV signal from the SM prediction, the injected signal can be
chosen to mimic the properties of the observed signal.)

Results should always be reported as bounds on $\sigma\times {\cal{B}}$ 
for any additional  $\phi$. It is reasonable to assume 
at this stage that its width, $\Gamma_\phi$, is small compared to the mass resolution. 
Nevertheless it is interesting and relevant to show the effect of varying this width. 
A useful way of presentation would be to summarize the information in ``temperature'' plots, such as Figure~\ref{fig:CMS_gamgam_like}, in the $(M_\phi, \Gamma_\phi / \Gamma^{\rm SM}_H)$ plane, with the 2D color map indicating the observed 95\% CL upper bound on the cross section. 
(Needless to say, the results should also be given in numerical form in addition to the plot.) 

Of course, specific model interpretations, for example bounds in the $\tan\beta$ versus $m_A$ 
plane in the MSSM,  are also interesting. However, they should be given only {\it in addition} 
to the $\sigma\times {\cal{B}}$ bounds.  
We stress this because it is important to be able to apply the observed
limits to general (beyond-) MSSM scenarios and notably to more general extensions
of the Higgs sector of the Standard Model. This is possible only if the
limits are presented in a less model dependent way, as outlined above.

\section{Fiducial cross sections}\label{sec:fiducial}

In situations in which the kinematic distribution of the signal depends on model parameters, 
simple scaling of production cross sections and decay branching ratios (relative to the SM) is not sufficient.  
Specifically, one must account for the change in the signal selection efficiency.
In order to address this broader class of theories, we advocate the measurement of fiducial cross sections for specific final states, {\it i.e.} cross sections, whether total or differential, for specific final states within the phase space defined by the experimental selection and acceptance cuts. 
This is meant in addition to, not instead of,  fits for signal strength modifiers $\mu$. 
Indeed, the (largely model-independent) fiducial cross sections and signal strengths w.r.t.\ SM 
are complementary to each other and both provide very valuable information in their own right.

With the full dataset of the LHC Run I, measurements of fiducial cross sections
with a precision of 20\% or so already become feasible in a number of channels.
In fact, ATLAS has already made the first attempt and released a preliminary fiducial
cross section for inclusive $H \to \gamma\gamma$ production:
$\sigma^{\mathrm{fid}} \times {\cal B} = 
56.2 \pm 10.5\,\mathrm{(stat)} \pm 6.5\,\mathrm{(syst)} \pm 2.0\,\mathrm{(lumi)}$~fb~\cite{ATLAS-CONF-2013-012}.
Fiducial cross section measurements require no model-dependent extrapolations to the full phase space, 
nor do they acquire additional theoretical uncertainty associated with such extrapolations. 
With carefully defined ``fiducial volumes'', the model-dependence of signal efficiencies
within such ``fiducial volumes'' can also be minimized so as to make it smaller than
the overall experimental uncertainties. For example, cuts on lepton transverse momenta
can be raised well above the knee of the efficiency plateau---this would minimize the impact
of possible variations in leptons' $p_T$-spectra on the overall signal efficiency.
Including isolation of leptons into the ``fiducial volume'' definition would help minimize
the sensitivity of a measured fiducial cross section on assumptions
about the jet activity in signal events. In some cases this is more difficult, for instance when the  the fiducial volume is defined by a cut on missing transverse energy,  which often introduces sensitivity to the topology of the event.  In situations where there is residual model-dependence in the fiducial efficiency, 
a service such as RECAST~\cite{RECAST} provided by the collaborations for explicitly calculating the 
fiducial efficiency would be of great value.

Fiducial cross sections, both total and differential, 
are standard measurements in high energy physics 
and for some processes are the only experimental cross sections available. 
For example, $J/\psi$ and $\Upsilon$ production cross section measurements at hadron colliders 
are always performed in some specified ``fiducial volumes''. This has allowed for a variety of models, 
many of which appeared or were substantially updated {\it after} the measurements had been made, 
to be confronted with the fixed experimental results.
In the context of Higgs boson physics, 
the fiducial cross sections can be categorized according to:
\begin{itemize} 

\item ``target'' decay mode, {\it e.g.}, 
$H \to ZZ \to 4\ell$, 
$H \to \gamma\gamma$, 
$H \to WW \to \ell\nu\ell\nu$, etc.; 

\item ``target'' production mechanism signatures, {\it e.g.}, 
(VBF-like $jj$)+$H$, 
$(\ell\ell)+H$, 
$(\ell+E_{T}^{\mathrm{mis}})+H$, 
$(E_{T}^{\mathrm{mis}})+H$, 
($V$-like $jj$)+$H$, etc.; 

\item and signal purity, {\it e.g.}, 
0-jet, 
1-jet,
high-mass VBF-like $jj$,
low-mass VBF-like $jj$, etc.

\end{itemize}

Fiducial cross sections can be interpreted in the context of whatever theoretical model, 
provided it is possible to compute its predictions for the fiducial cross section at hand 
({\it i.e.}, if it is possible to include experimental selection/cuts into the model). 
Typically, if the cuts defining ``fiducial volume'' can be implemented in a MC generator, 
this is rather straightforward. 
Therefore, complicated ``fiducial volume'' criteria ({\it e.g.} MVA-based) are not well suited, unless 
the MVA function is provided and depends only on kinematic information available at the generator level. 
Some reduction in signal sensitivity due to simplifications
in the event selection and due to possibly tighter cuts 
(to minimize the dependence of a signal efficiency on model assumptions as discussed above)
is an acceptable price. 

If these requirements for ``fiducial volume'' definitions are satisfied, 
then theoretical parameters of interest can be extracted from a fit 
to the measured cross sections.  As more than one fiducial cross section become available, 
to make a proper fit for parameters of interest, 
it is important that experiments provide a complete covariance matrix of uncertainties between
the measured fiducial cross sections. 

A parallel effort is required also from the theory community 
to develop the tools necessary for computing, with adequate precision, 
fiducial cross sections or ``fiducial volume'' acceptances 
with the associated uncertainties and their correlations
for the SM Higgs boson,
for a variety of BSM theories,
for an effective Lagrangian approach,
or for any other theoretical framework one might want to entertain. 

The ultimate measurements of an ``over-defined'' set of fiducial cross sections 
$\sigma^{\mathrm{fid}}_i$
can be unravelled into total cross sections associated with specific production mechanisms 
$\sigma^{\mathrm{tot}}_j$ 
via a fit of the following set of linear equations:
\begin{equation}
\label{eq:CrossSections}
\sigma^{\mathrm{fid}}_i = \sum_{j} A^{\mathrm{th}}_{ij} \times \sigma^{\mathrm{tot}}_j ,
\end{equation}
where $A^{\mathrm{th}}_{ij}$ are theoretical acceptances
of ``fiducial volumes'', in which 
fiducial cross sections $\sigma^{\mathrm{fid}}_i$ are measured.

The beauty of the concept of fiducial cross sections is that 
{\it experimental} uncertainties associated with 
measurements of fiducial cross sections $\sigma^{\mathrm{fid}}_i$ and
{\it theoretical} uncertainties associated with 
``fiducial volume'' acceptances $A^{\mathrm{th}}_{ij}$
are nicely factorized. Therefore, updates of theoretical acceptances/uncertainties
or a confrontation of emerging new models with experimental results
do not require a re-analysis of experimental data. 
One can also treat the total cross sections $\sigma^{\mathrm{tot}}_j$ 
as nuisance parameters and fit data for 
theoretical acceptances $A^{\mathrm{th}}_{ij}$ 
({\it e.g.}, a 0-jet veto acceptance), 
if it is these quantities that one is primarily interested in.

We would like to advocate that experiments do measure fiducial cross sections 
even at 8~TeV in as many final states as feasible, however small this number might be. 
The future LHC center-of-mass energies will be higher and no more updates
for the 8 TeV fiducial cross sections will be likely. 

Finally, we note that measurements of differential fiducial cross sections, when they become possible, 
will be even more powerful (in comparison to just total exclusive fiducial cross sections)
for scrutinizing the SM Lagrangian structure of the Higgs boson interactions, including 
tests for new tensorial couplings, 
non-standard production modes, 
determination of effective form factors, etc.

\clearpage
\section{Executive summary}

With the LHC operations at 7--8~TeV in 2010--2012, we 
have just begun the exciting exploration of the TeV scale. 
Natural stabilization of the EW requires new phenomena at TeV energies---the 
measurements of the Higgs properties may provide a guide as to where and how to look 
for this New Physics. Moreover, if New Physics is discovered, combining it with the 
results from the Higgs sector will be essential for establishing the underlying 
fundamental theory beyond the SM. 
It is therefore of utmost importance that the Higgs results 
be usable by the whole high-energy-physics community.
To this end, we put forth the following suggestions regarding the presentation of the Higgs results:

\begin{itemize}

\item For each Higgs decay mode $Y$ ($\gam\gam$, $WW$, $ZZ$, $b\bar b$, $\tau\tau$ are currently considered) provide the likelihood ${\cal L}$ of the signal strengths in the $(\mu(X,Y), \mu(X',Y))$ plane, as shown in Figure~\ref{fig:CMS_gamgam_like}. The grouping $X=\mathrm{ggF+ttH}$, $X'=\mathrm{VBF+VH}$ is well motivated, but additional choices of $X$ and $X'$ should be considered when appropriate for the given analysis.  
The content of the plots should {\em always} be provided also in numerical form, {\it e.g.}, as a {\tt ROOT} file or as a simple text file with a grid. 
In addition to the combined results, results should also be given separately for each 
$\sqrt{s}$. 

\item To go a step further and overcome the limitations induced by 2D projections and/or combining production modes, 
provide the signal strength likelihood as a function of $m_H$, separated into all five production modes 
ggF, ttH, VBF, ZH and WH; {\it i.e.}\ for each decay mode considered give the likelihood in the 6D form 
${\cal L}(m_H, \mu_{\rm ggF},\mu_{\rm ttH},\mu_{\rm VBF},\mu_{\rm ZH},\mu_{\rm WH})$. 
Ideally, this should again also be done separately for each $\sqrt{s}$. 

\item Concerning searches for additional Higgs-like states with masses above
or below 125 GeV, provide the results including the injection of a signal 
with the properties of a SM-like Higgs boson at 125--126 GeV. 
Moreover, always present the results as bounds on pure $(\sigma\times \br)$ in addition to any model interpretation. 

\item Whenever possible, provide kinematic event selection criteria 
that can approximately be reproduced by phenomenologists,  
{\it e.g.}, using Monte Carlo event generators.\footnote{It is understood that 
this typically requires simplified versions of the analyses, which are sub-optimal. However, apart from measurements of form factors as outlined in Section~\ref{sec:tensor}, this provides at present the only practical means for BSM interpretations of the experimental results involving models with altered matrix element structures (leading to altered kinematical distributions) or additional production modes.} 
The desired information is: the complete cut flow, estimated
number of background events, expected event yields for all the SM Higgs processes,
and the observed number of signal events or limits thereon.
For MVA-based analyses, it would be of great value if a simplified version of the final MVA could be given. 

\item In addition to direct model-dependent interpretations of data,
the long-term goal should be to develop a consistent scheme for publishing 
fiducial cross sections $(\sigma^{\mathrm{fid}} \times \br )$, either
measurements or limits for null search results,
as done conventionally for SM processes. 

\item We suggest that this supplementary material is made available via INSPIRE~\cite{inspire}. This way the complete set of information will be searchable, citable, and accessible from a single point.

\end{itemize}
By following these procedures, the existing data will be of maximal utility to the whole high-energy-physics community  in assessing new models and scenarios for the Higgs sector. 
We note that INSPIRE is a natural platform in our field to make available such additional material that cannot reasonably be included within the traditional text publications. In particular, INSPIRE also allows one to associate to each article auxiliary information which maximize the utility of the data such as electronic form of plots and multi-dimensional information. Indeed, it will be of great advantage if, wherever possible, results are given in multi-dimensional form, not just projected onto 2D planes. 
We note that it is already common practice in the LHC experiments to provide useful 
auxiliary information on HepData~\cite{hepdata}, in RIVET~\cite{Buckley:2010ar},  
and/or on collaboration twiki pages.  
The INSPIRE project may help to build a coherent information system.\footnote{Supplementary data submitted to HepData is automatically made available and fully searchable on INSPIRE.} 
In particular, INSPIRE is able~\cite{sunje} to assign Digital Object Identifiers (DOI)~\cite{doi} to this auxiliary information.  This persistent identifier ensures that these supplementary materials are uniquely identifiable, searchable and citable. 

\appendix
\section{Appendix}\label{A:appendix}
In September 2013, after the original submission of this note, the ATLAS collaboration digitally published the likelihood in a grid on the $(\mu_{\rm ggF + ttH}, \mu_{\rm VBF + VH})$ plane associated to \\$H\to\gam\gam$, $H\to ZZ^* \to 4\ell$, and $H\to WW^* \to \ell\nu\ell\nu$~\cite{ATLAS-data-Hgamgam,ATLAS-data-HZZ,ATLAS-data-HWW}.


\end{document}